\documentclass[aps,pre,amsmath,amssymb,
reprint,superscriptaddress]{revtex4-2}
\usepackage{enumerate}
\usepackage{amsmath}
\usepackage{amsthm}
\usepackage{amssymb}
\usepackage{graphicx}
\usepackage{floatflt}
\usepackage{fancybox}
\usepackage{appendix}
\usepackage{enumitem}
\usepackage{subfigure}
\usepackage{array}
\usepackage{dsfont} 
  \usepackage{placeins}
\usepackage{lipsum}  
\usepackage{color}  
\usepackage[normalem]{ulem}
\usepackage{hyperref}
\usepackage[b]{esvect}
\usepackage{afterpage}
\usepackage{tikz}
\usetikzlibrary{tikzmark}
\UseRawInputEncoding

\usepackage{color}

\renewcommand{\Re}{\textrm{Re}}
\renewcommand{\Im}{\textrm{Im}}

\renewcommand{\vec}[1]{\mathbf{#1}}
\newcommand{\p}{\partial}

\newcommand{\be} {\begin{equation}}
\newcommand{\ee} {\end{equation}}
\newcommand{\bsub}{\begin{subequations}}
\newcommand{\esub}{\end{subequations}}
\newcommand{\bea}{\begin{eqnarray}}
\newcommand{\eea}{\end{eqnarray}}
\newcommand{\bi} {\begin{itemize}}
\newcommand{\ei} {\end{itemize}}
\newcommand{\ben} {\begin{enumerate}}
\newcommand{\een} {\end{enumerate}}
\newcommand{\bmat} {\begin{pmatrix}}
\newcommand{\emat} {\end{pmatrix}}
\newcommand{\bal} {\begin{aligned}}
\newcommand{\eal} {\end{aligned}}
\newcommand{\btab}{\begin{tabular}}
\newcommand{\etab}{\end{tabular}}

\begin{document}

\date{\today}

\title{Exact analogue of the Hatano-Nelson model in 1D continuous nonreciprocal systems}

\author{A. Maddi}

 \address{Laboratoire dÕAcoustique de lÕUniversitŽ du Mans (LAUM), UMR 6613,  Institut dÕAcoustique - Graduate School (IA-GS), CNRS, Le Mans UniversitŽ, France.}

 \author{Y. Auregan}%
 \address{Laboratoire dÕAcoustique de lÕUniversitŽ du Mans (LAUM), UMR 6613,  Institut dÕAcoustique - Graduate School (IA-GS), CNRS, Le Mans UniversitŽ, France.}

 \author{G. Penelet}%
 \address{Laboratoire dÕAcoustique de lÕUniversitŽ du Mans (LAUM), UMR 6613,  Institut dÕAcoustique - Graduate School (IA-GS), CNRS, Le Mans UniversitŽ, France.}
 \author{V. Pagneux}%
 \address{Laboratoire dÕAcoustique de lÕUniversitŽ du Mans (LAUM), UMR 6613,  Institut dÕAcoustique - Graduate School (IA-GS), CNRS, Le Mans UniversitŽ, France.}
 
 \author{V. Achilleos}%
 \address{Laboratoire dÕAcoustique de lÕUniversitŽ du Mans (LAUM), UMR 6613,  Institut dÕAcoustique - Graduate School (IA-GS), CNRS, Le Mans UniversitŽ, France.}

\begin{abstract}
% The field of non-Hermitian physics and especially  nonreciprocal systems have 
% underlying fascinating properties and potential applications. A prototype system to this extend is the Hatano-Nelson (HN) model which is a nonreciprocal discrete model. 
% We propose a general model which allows for  continuous nonreciprocal periodic systems in 1D to be exactly mapped  onto the Hatano-Nelson model. Our broadband approach relies on the two port transfer matrix and thus is applicable to a plethora of physical systems. An example of  a physical implementation of our model is considered in acoustic waveguides. We show both theoretically and experimentally that using active acoustic elements we achieve the mapping to the HN model and observe the celebrated skin effect. Furthermore, our setup allows us to investigate experimentally the transition from a periodic boundary condition  to an open boundary condition using diaphragms of different radii. This way, we uncover the exponential sensitivity of the system to boundary condition changes. Our work provides a premier mapping of continuous systems to well known nonreciprocal discrete models and paves the way into extending non-Hermitian and topological properties in experimentally feasible.
We propose a general framework  that enables the exact mapping of continuous nonreciprocal 1D periodic systems to the Hatano-Nelson (HN) model. Our approach, based on the two-port transfer matrix, is broadband and is applicable across various physical systems and, as an illustration, we consider the implementation of our model in acoustic waveguides. Through theoretical analysis and experimental demonstrations, we successfully achieve the mapping to the HN model by utilizing active acoustic elements, thereby observing the renowned skin effect. Moreover, our experimental setup enables the exploration of the transition from periodic to open boundary conditions by employing diaphragms of varying radii. Our experimental results, unveil the exponential sensitivity of the system to changes in boundary conditions. 
By establishing a profound connection between continuous systems and the fundamental discrete HN model, our results significantly broaden the potential application of nonreciprocal wave systems and the underlying phenomena.
\end{abstract}
\maketitle

%\tableofcontents

\section{Intro}
%
%I. General nonhermitian: from PT to the new more generalised models \\
%II. Skin effect in various setups\\
%III. Experiments/theory and where our work is different\\

In recent years, the interest in the intriguing features of non-Hermitian Hamiltonians has increased extensively\cite{ashida2020non,el2018non}. These Hamiltonians enable the study of non-conservative systems, characterized by complex eigenenergies, that reflect the presence of gains/losses. The growing attention to such systems, despite their inherent complexity,  builds upon the pioneering work of Bender  on PT symmetry \cite{bender1998real,bender2002generalized}, who first demonstrated through his study of parity and time conserved operator that non-Hermitian systems can still exhibit real eigenenergies when gains and losses are perfectly balanced. Later on, many novel features of PT-symmetric systems were uncovered, such as particular sensitivity\cite{liu2016metrology,miri2019exceptional,wiersig2020review,chen2017exceptional}, CPA-Lasing \cite{auregan2017p,longhi2010pt,chong2011p,poignand2021parity,wong2016lasing} and unidirectional invisibility \cite{lin2011unidirectional,mostafazadeh2013invisibility}.

More recently, there has been a surge of interest in the interplay of non-Hermiticity and topological phenomena given their promise for the unidirectional control of waves \cite{wanjura2021correspondence,longhi2022self,weidemann2020topological},  and the development of zerod sensors \cite{edvardsson2022sensitivity,budich2020non,mcdonald2020exponentially,koch2022quantum}. In this perspective, the topological phase of matter explores the relationship between the bulk properties of a lattice and its behavior at the boundary, using topological invariants. This has led to the discovery of new non-Hermitian properties, such as the Non-Hermitian skin effect\cite{okuma2020topological,zhang2022review}, which occurs when transitioning from periodic boundary conditions (PBC) to open boundary conditions (OBC) and results in the localization of energy at one boundary. This effect has been extensively studied theoretically  with experimental demonstration in electrical circuits\cite{xu2021coexistence,helbig2020generalized,liu2021non} and acoustic \cite{zhang2021acoustic}.

The Hatano-Nelson \cite{hatano1996localization} (HN) model is one of the most prominent models in the field of non-Hermitian topology. This model describes a one-dimensional lattice where each site is coupled by a pair of asymmetric (nonreciprocal) hopping. This asymmetry shifts the bulk states towards the boundaries, localizing them at the side of the stronger hopping, resulting in a non-Hermitian skin effect (NHSE). 
In addition, since the recent work of Ref.\cite{Wangtopononher2019} the topological properties of the model and of its generalizations have given birth to a  field of no-Hermitian topology in discrete lattices\cite{reviewBerg,reviewSato,reviewgood}.
Since the breaking of reciprocity is a prerequisite to observe the NHSE, the experimental realization of such systems is rather challenging, as it typically requires an external energy source/sink, and can potentially lead to instabilities.  

Taking a step forward, in this study, we propose a comprehensive, broadband and exact mapping of the HN model to \textit{continuous} 1D nonreciprocal periodic systems. The approach adopted consists in the development of a general theoretical framework that allows the mapping of a 1D linear nonreciprocal system described by its unit cell transfer matrix.  We validate our model using the example of an acoustic waveguide and furthermore we performed experiments employing a network of active loudspeakers \cite{penelet2021broadband} to observe the NHSE. Our setup allows us to also build a stable setup with periodic boundaries, due to the inherent losses of the system. Using diaphragms we experimentally observe the transition of the eigenfrequencies from periodic boundary conditions (PBCs) to open boundary conditions (OBCs) and exhibit the exponential sensitivity of the system to changes of the boundaries.
%.  and used a diaphragm of varying radii to demonstrate the experimental transition of eigenfrequencies of the acoustic problem from PBC to OBC, as well as the intermediate stages. This work provides(... bla bla)

\section{Model}
\subsection{Hatano-Nelson model for 1D continuous systems}

 \begin{figure}[ht]
\includegraphics[width=8.6cm]{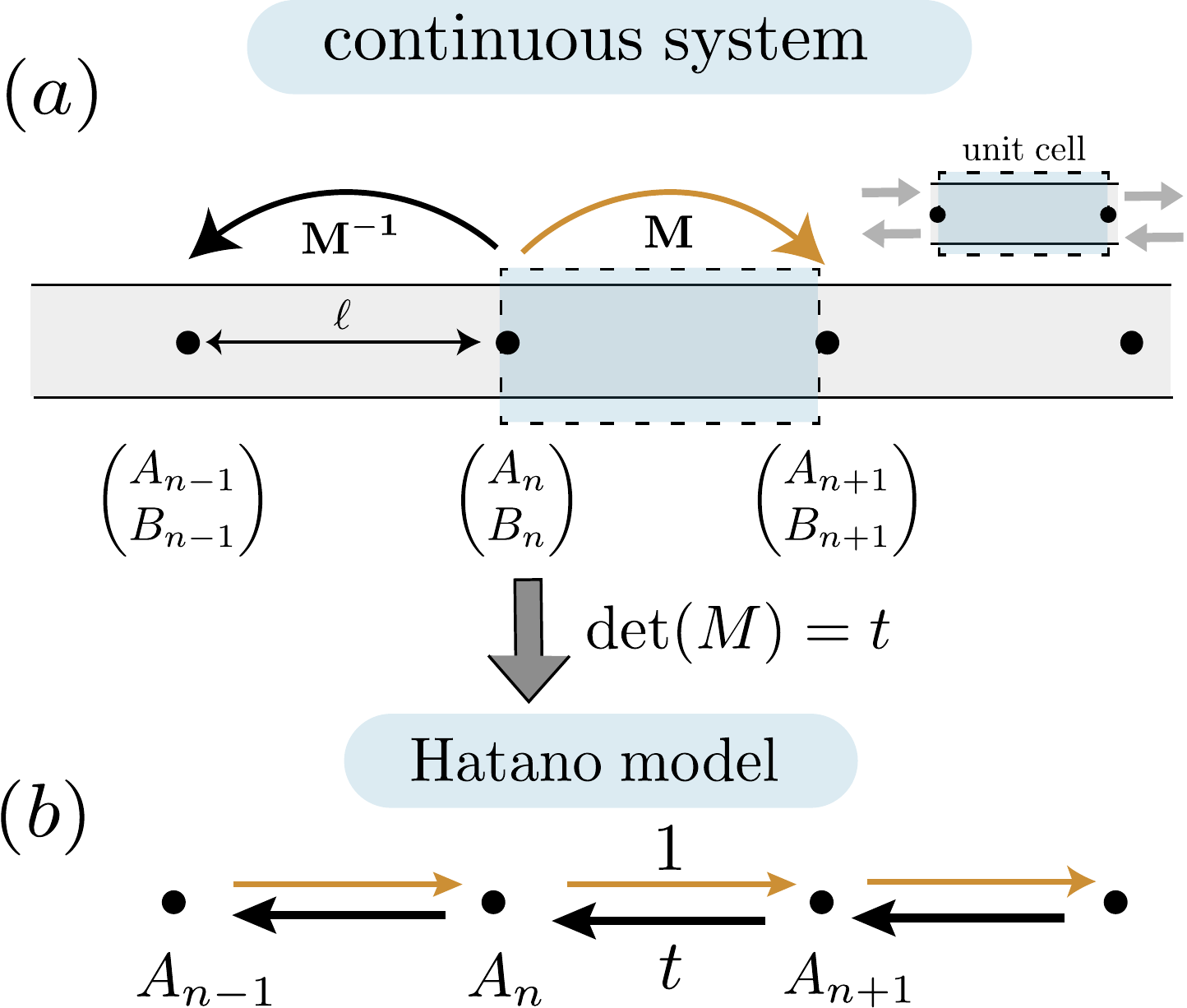}\\
\caption{(a) A sketch of a continuous system which is composed by periodically arranged unit cells of length $l$.
The edges of each unit cell are connected via the transfer matrix $\mathbf{M}$. (b) By a simple manipulation of the transfer matrix equations we map the continuous system to the HN model where the nonreciprocal coupling strength $t$ is equal to the determinant of the transfer matrix. }
\label{fig1}
\end{figure}

We consider wave propagation in a continuous medium which is periodic and at the edges of its unit cell only one mode is propagating (monomode approximation). Such a 2-port unit cell can be described using a two by two transfer matrix of the general form [see also Fig.\ref{fig1} (a)]

%More particularly, the transfer matrix of the unit cell has the following general form  
\begin{align} 
\mathbf{M} \color{black}=\begin{pmatrix}  a & b\\ 
c & d \end{pmatrix}, \quad \rm{det}(\mathbf{M} \color{black})=t.
\label{M}
\end{align}
%
%and connects two points of the system as shown in  Fig.\ref{fig1} (a). 
%As long as the system is non-reciprocal then $t\ne 1$.
The state vector of the system at the position $x$ takes the form $[A(x), B(x)]^T$ and for simplicity below we use the notation $A_n\equiv A(x_n)$. 
 Eq.(\ref{M}), allows us to write the following system of equations between three consecutive equidistant points, \color{black}
\begin{align} 
\begin{pmatrix}A_{n+1}  \\ B_{n+1} \end{pmatrix} =\mathbf{M} \begin{pmatrix} A_{n}  \\ B_{n}\end{pmatrix},\;
\begin{pmatrix}A_{n-1}  \\ B_{n-1}\end{pmatrix} =\mathbf{M}^{-1} \begin{pmatrix} A_{n}  \\ B_{n} \end{pmatrix}. 
\label{transf1}
\end{align}
The first line of each of the above system of equations is explicitly written as
\begin{align} 
A_{n+1}&=a A_{n}+ b B_{n}, \label{transf21} \\ 
tA_{n-1}&=d A_{n}- b B_{n}.\label{transf22} 
\end{align}
%
%We can thus add the two Eqs.(\ref{transf21}) and (\ref{transf22}), to obtain
 %the following simple relation for the acoustic flux
 Therefore, by adding the two Eqs.(\ref{transf21}) and (\ref{transf22}), the problem can be simplified to the following discrete equation \color{black}
\begin{align} 
A_{n+1}+t A_{n-1}=E A_n.
\label{HN}
\end{align}
%
%where we have used the notation $u_n\equiv u(x_n)$.
The last equation provides an exact analogue of the Hatano-Nelson (HN) model and can be readily applied to 
various continuous physical systems. Note that the same equation is also true for $B_n$.
%which exactly the HN model. Here we have used the notation $u_n \equiv u(x_n)$. and the nonreciprocal hopping is simply given by the determinant of the transfer matrix $t$.
According to our mapping the transfer matrices of both the continuous and the discrete unit cell have the same non unitary determinant. This is important since it was recently shown in \cite{kunst2019non} that the nonzero determinant is a key parameter to re-establish the bulk-boundary correspondence for non-Hermitian systems.
The \textit{energy} $E$ in Eq.\ref{HN} is simply given by
\begin{align} 
E=\rm{tr}(\mathbf{M})=a+d,
\label{map0}
\end{align}
and provides the direct link between the eigenvalues of the Eq.\ref{HN} and the  elements of $\mathbf{M}$.
%frequency $\omega=k/c_0$. 
In practice, for wave systems the only restriction of our model is that the determinant $t$ of the transfer matrix does not depend on the frequency. When the latter is satisfied our exact analogue of the HN is broadband and only requires periodicity, the monomode approximation and a non-unimodular transfer matrix. 
\color{black}
%

%
% \begin{align} 
%E(k)=(t+1)\cos(kl)-Z_0^{-1}\sin(kl)\left(X +i\beta\right).
%\label{map}
%\end{align}

%

%\color{red}
%Note that the introduced exact analogue of the HN model is broadband and only requires periodicity and the plane wave approximation. The form of the mapping $E(k)$ is given in the Appendix for the case of a unit cell which has a loudpseaker with a feedback loop at its center. The mapping includes both the effects of the feedback loop and also the losses in the system. 

\begin{figure}[h!]
\includegraphics[width=8.6cm]{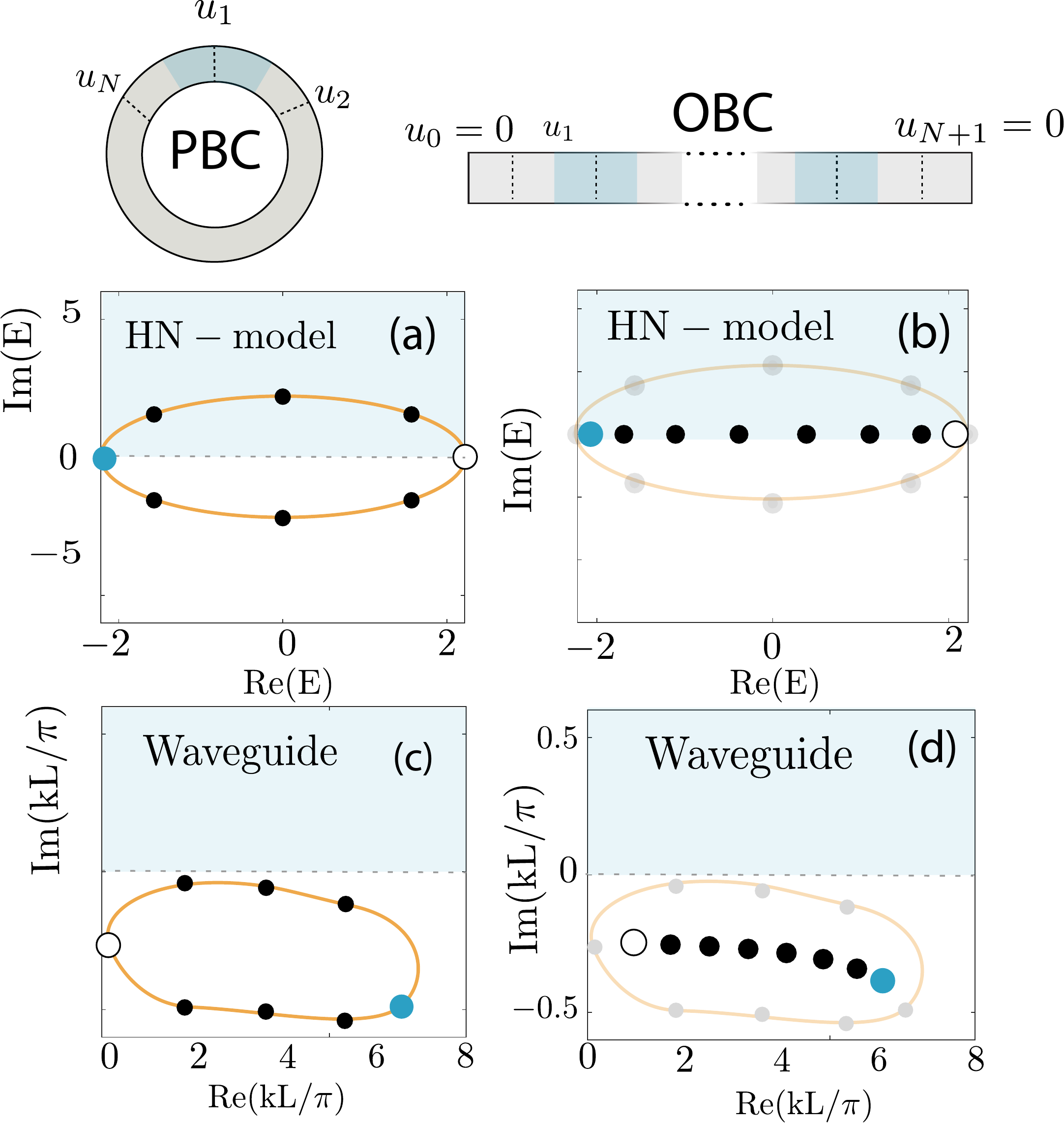}\\
\caption{(a)
 (a-b) The energy of the modes in the complex plane for a periodic HN lattice with periodic (PBC) and open (OBC) boundaries.
  (c-d) The corresponding acoustic complex frequencies of the waveguide. In all panels, the number of sites is set to $N=8$ and the hopping factor to $t=1.2$\color{black}.  }
\label{fig2}
\end{figure}

We now briefly summarize the properties of the HN model %regarding its eigenmodes and eigenfrequencies $E$ and 
%and illustrate how these are translated into the acoustic problem.
%First of all considering Bloch wave solutions of Eq.(\ref{HN})
%in the form $u_n=U\exp(-iqn)$ we find the dispersion relation of an infinite system to be 
%In this section, a brief summary the HN model properties is proposed.
starting from the dispersion relation obtained by considering periodic solutions of Eq.(\ref{HN}) in the form $A_n=A\exp(-iqn)$ which yields
%, anthe following dispersion relation of an infinite system is found, 
%
\be
E_q=(1+t)\cos(q)+i(1-t)\sin(q).
\label{HNdr}
\ee
When $t\ne 1$, in the presence of non-reciprocity, the energies are complex. Furthermore $E(q)$ creates a closed loop in the complex plane for $q\in [-\pi/2,\pi/2]$. The fact that the energy itself is a complex function has motivated researchers to attribute topological  properties to such non-Hermitian models. In particular it is now well established that one can define the following winding number
\be
w_E=\frac{1}{2\pi}\oint_\mathcal{C}\frac{d}{dz}\rm{arg}E(z),
\label{winding}
\ee
where $z=\exp(iq)$. This  integral along the Brillouin zone gives $w_E=1$ ($w_E=-1$) for $\vert t\vert>1$ ($\vert t\vert<1$) signaling a transition at $=0$. This transition is now known to be related with the appearance of the so called skin modes\cite{zhang2020correspondence}, i.e. localised modes at one edge of a finite structure. The sign of the winding number indicates (in 1D) the side at which the skin modes are localised. 
%These results have been generalised for various models including next-neighbour interactions and different geometries.
%\subsection{HN eigenvalues and the corresponding  acoustic modes}
%
Let us now make a quick remainder of the spectra of the system under PBC and OBC.
\paragraph{PBC} %We now discuss the case of a finite lattice with periodic boundary conditions.
%For finite lattices, the appearance of skin modes depending on the boundary conditions is easily understood by studying the underlying eigenvalue problems.
For the case of a lattice of $N$ sites with periodic boundary conditions the corresponding eigenvalue problem is
\be
H_{\mathrm{PBC}}\mathbf{A}=E\mathbf{A}, \quad H_{\mathrm{PBC}}= \bmat 0 & 1 & 0 & \dots & t \\ 
t & 0 & 1 & & \vdots \\ 
0 & t & 0 & \ddots & 0  \\ 
\vdots & & \ddots & \ddots & 1 \\ 
1 & \dots & 0 & t & 0 \emat,
\label{Hper}
\ee
where $H_{\mathrm{PBC}}$ is a circulant matrix  and its eigenvalues are given by Eq.\ref{HNdr} after replacing $q_n=2\pi n/N$.
The corresponding eigenvectors are known to be of the following form
\be
A_j =\frac{1}{N} \bmat 1,\lambda^j,\lambda^{2j} \dots \lambda^{(N-1)j}  \emat^T,
\ee
where $\lambda=e^{\frac{i2\pi}{N}}$ is the N-th root of unity. These modes are independent of $\vert t\vert$ and are thus extended.
Regarding the spectrum, there is a big difference between the case with $\vert t\vert=1$ and the one with $\vert t\vert\ne 1$. The spectrum of the lattice model undergoes a change from a straight line into a closed loop in the 
 complex. An example for $t=1.2$ and $N=8$ is shown Fig\ref{fig2} (a). 
\paragraph{OBC} 
The next most studied scenario is the case of open boundary conditions i.e. when $A_0=A_N=0$. For the  continuous system this translates to Dirichlet boundary condition of $A(x)$ at the ends. The corresponding eigenvalue problem can be written as  
\be
H_{\rm{OBC}} \vec{A}=E \vec{A},\quad
H_{\rm{OBC}}= \bmat 0 & 1 & 0 & \dots & 0 \\ 
t & 0 & 1 & & \vdots \\ 
0 & t & 0 & \ddots & 0  \\ 
\vdots & & \ddots & \ddots & 1 \\ 
0 & \dots & 0 & t & 0 \emat .
\label{OBC}
\ee
The latter matrix is tridiagonal Toeplitz, with positive off-diagonal elements. Interestingly, although $H_{\rm{OBC}}$ is non-symmetric it can be transformed into a \textit{symmetric} matrix under the similarity transformation $\tilde{H}=D^{-1}H_{\rm{OBC}}D$ where $D=\rm{diag}(d_0,d_1..d_{N-1})$ with elements $d_n=\sqrt{t^n},\quad n=0,\ldots N-1$. The eigenvalues of $H_{\rm{OBC}}$ are real and are given by the following expression
\be
E_n=2\sqrt{t}\cos(\frac{n\pi}{N+1}),\quad n=1,\ldots, N.
\label{eigsfixed}
\ee
In addition the  $j$-th  right eigenvector of  $H_{\rm{OBC}}$ has the following form
\be
%A^R_j=\bmat A^R_{j,1}, A^R_{j,2} \cdots A^R_{j,N}\emat^T, \; A^R_{j,n}= (t)^{n/2}\sin\frac{jn\pi}{N+1}.\\
A^L_j=\bmat A^L_{j,1}, A^L_{j,2} \cdots A^L_{j,N}\emat^T, \; A^L_{j,n}= \color{black}t^{n/2}\color{black}\sin\frac{jn\pi}{N+1}.
\label{eigvobc}
\ee
%
%The notation $u^R$ is to remind that right eigenvectors are not the same as the left ones.
 It is clear that due the prefactor $(t)^{n/2}$ \color{black} (which is independent of the eigenvalue $j$), as long as  $t> 1$ ($t<1$) \color{black} all states localize on the right (left) hand side. This is exactly the skin effect used to describe the fact that all modes are localized at one edge.  

\vspace{10pt}
Note that all the results derived apply to $B_n$ as well and if the  necessary boundary conditions (i.e $B_0=B_N=0$) are assigned we derive the same results.\color{black}
% From now on we focus on waves at single frequencies $\omega=c_0k$ where $k$ is the wavenumber, $c_0$ the speed of sound
% and we will use the time convention of $e^{-i\omega t}$. The system under consideration is illustrated in Fig.\ref{fig1} (a) and is composed by periodically arranged unit cells of length $\ell$. 
% Under this assumptions one can describe wave propagation using the transfer matrix which relates the acoustic pressure  $p$ and the flux velocity $u$ . 
\subsection{An example: nonrecirocal periodic acoustic waveguides}

According to our proposed model, we \sout{thus} expect that both the closed loop spectrum and the skin modes of the HN model can be observed in a continuous medium.
We now give an example of a physical system that can be mapped to the HN model following the aforementioned procedure. Nonreciprocity in acoustics can be achieved using (among others) active elements, spatio-temporal modulations, or the thermoacoustic effect \cite{nassar2020nonreciprocity,olivier2022asymmetric,penelet2021broadband,fleury2014sound,popa2014non,zhai2019active,geib2021tunable}. Here we will use the idea of an active loudspeaker to break the reciprocity and achieve $\vert t\vert\ne 1$. To find the corresponding acoustic modes, we need the elements of the matrix $M$ for a particular setup. Here, we use the transfer matrix corresponding to a unit cell of length $l$ and a loudspeaker with a feedback loop mounted in the middle of the cell. In view of the experiments in acoustics the appropriate transfer matrix is the one connecting the acoustic pressure $p(x)$ and the acoustic flux $u(x)$ thus identifying $A\rightarrow u$ and  $B\rightarrow p$. For low frequencies where the monomode approximation is valid we may then write an anlytical expression for the transfer matrix elements which leads to the following mapping (see Appendix)
%\begin{align} 
$E(k)=(t+1)\cos(kL)-\left(a\frac{k^2-k_0^2}{k}+i\beta(k)\right)\sin(kL)$.
%\label{map}
%\end{align}
%
Here we have used the notation for the frequency $\omega=c_0k$ with $k$ the wavenumber and
$c_0$ the speed of sound.
% and we will use the time convention of $e^{-i\omega t}$.
Furthermore, $k_0$ and $\beta$ denote the resonance frequency of the loudspeaker and its electro-mechanical losses respectively. Using the expression for $E(k)$ each eigenvalue for the PBC [Eq.(\ref{HNdr})] or the OBC [Eq.(\ref{eigsfixed})] is mapped to the corresponding acoustic eigenfrequency $k_n$. One such example of eigenfrequencies is plotted in panels (c) and (d) of Fig\ref{fig2} for lattice with $t= 1.2$.

An important observation here is that for the PBC, although the HN model predicts a set of generally unstable modes (with $\Im(E)>0$), the losses of the acoustic system, embedded in the mapping, allows for a closed loop spectrum below the real axis and thus stable acoustic modes. In accordance, the straight line spectrum of the OBC for the acoustic system becomes an arc lying in the stable part of the complex plane.
Another consequence of the mapping is that the smallest (largest) eigenvalues are swapped (see white and blue points the panels of Fig\ref{fig2}).

\section{Experimental realization and the transition from PBC to OBC}

\subsection{Skin effect and measured spectra}

\begin{figure}[h]
    \centering
    \includegraphics[width=86mm]{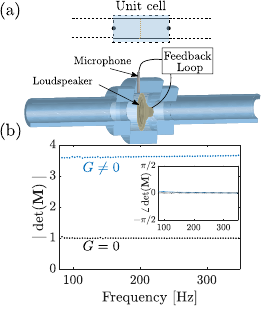}
    \caption{\textbf{Experimental unit cell.} (a) sketch of the unit cell (b) determinant of the transfer matrix of the unit cell (equivalently the hopping factor $t$) as a function of the frequency, where black and blue represents respectively the passive and active cell.   }
    \label{detT}
\end{figure}

We now  show the experimental realization of the acoustic HN model. A sketch of the unit cell used used in the experiment is displayed in Figure\ref{detT} (a) and consists of a cavity connected to two ducts. A speaker is installed in the center of the cavity and is controlled by a feedback loop consisting of a current amplifier and a microphone mounted in the vicinity of the loudspeaker. 
%It should be noted that the cross sectional change (i.e the cavity) was introduced here only to reduce the loss of the resonant system (i.e by reducing the volume flow passing through the loudspeaker). 
The non-reciprocity arises from the electroacoustic feedback loop, in which an electrical current supplied to the loudspeaker is proportional to the feedback gain $G$ and the pressure measured by a nearby microphone. This generates an additional oscillating force that acts on the loudspeaker membrane. For frequencies below the cutoff, the acoustic pressure and velocity at the edges of the unit cell are connected through a transfer matrix $\mathbf{M}$ as in Eq.(\ref{M}), and the hopping parameter $t$ is simply adjusted by the amplifier gain $G$. %For a more detailed derivation of the transfer matrix, please refer to the supplementary material.

To confirm the nonreciprocity of the unit cell, the experimentally measured determinant of the transfer matrix $\text{det}(\mathbf{M})$ is displayed as a function of the frequency in Figure\ref{detT}.(b). This measurement was conducted using an impedance sensor \cite{macaluso2011trumpet}, further details can be found in the supplementary information. In the absence of a feedback loop $G=0$, the system is reciprocal (i.e., $t=\text{det}(\mathbf{M})=1$) since the loudspeaker behaves as a passive resonator. However, if a gain $G\ne 0$ is applied, the reciprocity is broken and the hopping term $t$ becomes nonunitary, thereby favoring propagation in one direction. For the measurements shown in Fig.\ref{detT}.(b) we have tuned the gain such that $t=3.7$ which would lead to a right-side localization with ($w_E>1$). Furthermore, one can see that the hopping factor is independent of the frequency and thus the mapping is broadband. 

\begin{figure}[ht]
    \centering \includegraphics[width=86mm]{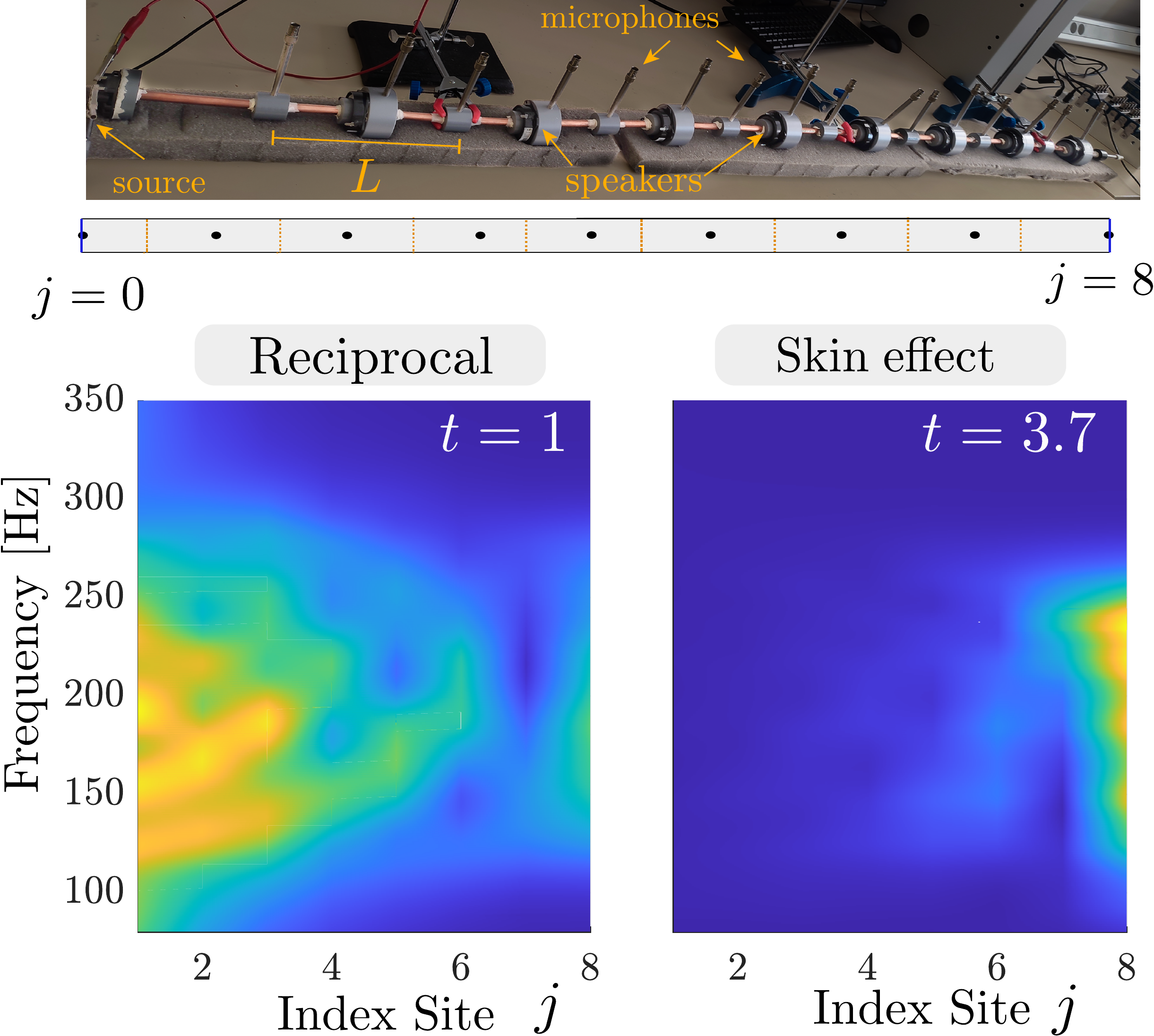} \caption{\textbf{OBC configuration}. The experimentally measured pressure as a function of  the index site $j$ and the frequency $f$, for a symmetrical ($t=1$, left) and asymmetrical ($t=3.7$, right) hopping, where the system is excited from the left side at $j=0$.}
    \label{OBCsys}
\end{figure}

For the experimental realization of the HN model with OBC and the observation of skin-modes, a periodic system composed of $N=8$ identical unit cells is constructed. The two ends of the waveguide are then closed with rigid walls at the extremities which corresponds to Dirichlet boundary conditions for the acoustic flux. We excite the system from the one end (left) and measure the pressure at equidistant points designated by $p(\omega,x_j)$ as shown in the top of Fig.\ref{OBCsys}. 

%Two different kind of boundaries  will be imposed, the $\textit{OBC}$ corresponds to rigid walls at extremities of the waveguide whereas the $\textit{PBC}$ corresponds to looped configuration. The system is then excited using a monochromatic source at a selected position and the acoustic pressure is measured at the edges of each unit cell designated by $p(\omega,x_j)$. 

%
%at both ends of each cell is measured as well as the pressure of the feedback microphone, but for ease of reading, only those at the extremities of the cells will be presented and 

\begin{figure*}[ht] 
    \centering
    \includegraphics[width=129mm]{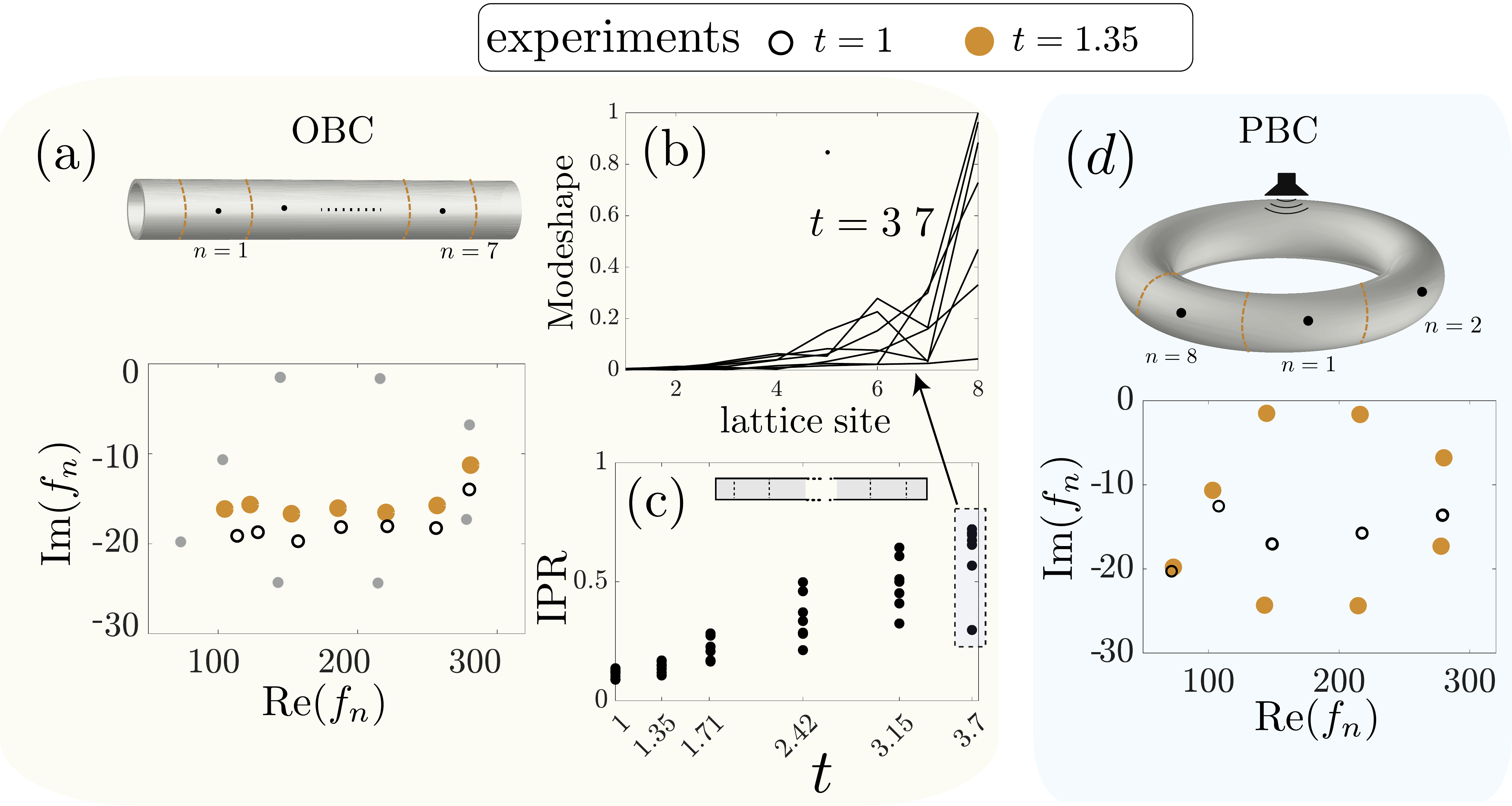}
    \caption{\textbf{Experimental results.}(a) complex eigenfrequencies of the problem in the PBC and OBC configuration, for $t=1$ (white) and $t=1.35$ (orange). (b) The inverse participation ratio of each modeshapes as a function of the hopping factor $t$ in the OBC configuration. (c) Modeshape of the OBC configuration for $t=3.7$.}
    \label{fig5}
\end{figure*}

The bottom panels of Fig.\ref{OBCsys} depict the magnitude of the measured acoustic pressure at different sites as a function of  frequency.
%the system has hard wall at the edges and is excited using a loudspeaker placed at the location $x_{j=0}$. 
Starting with the reciprocal case ($t=1$), the system is excited and the propagation occurs only within the interval $f\in [120,300]$ Hz which is the first allowed band of the periodic lattice.
Furthermore, the field appears to have greater amplitude near the source due to the damping (mainly caused by the loudspeakers), since away from the source the wave is rapidly dissipated. %which causes the energy to be spread throughout the system but with a greater concentration at the proximity of the acoustic source. 

On the other hand, by turning on the feedback gain and reaching a value of the asymmetric hopping $t=3.7$, we clearly see the appearance of the non-hermitian skin effect at the opposite boundary of the system. This means that despite the high damping, any excitation from the left leads to a strong localization on the right side, with an amplification ratio from the first site $j=1$  to the last  $j=N$ roughly equal to $p(x_N)/p(x_1)=120$. 
Note that the accumulation of energy on the right hand side is persistent for all frequencies in this band 
confirming the fact that all modes exhibit the skin effect.
This amplification is in quantitative agreement with the OBC solutions of Eq.(\ref{eigvobc}) where the amplification ratio for the modes is $\sim t^{N/2}$.

Now, we focus in more detail to evaluate the experimentally obtained complex eigenfrequencies $f_n$ of the corresponding modes of the underlying cavity. These can be estimated using different fitting algorithms in the framework of the so called experimental modal analysis \cite{allen2006global,peeters2004polymax}. Such algorithms have proven to be reliable in the study of complex structures. To operate, they require a collection of frequency response functions FRFs. Herein, it takes the form of the measured acoustic pressure $p(\omega,x_j)$. 
 In addition, the proposed acoustic model allows us to perform such measurements also using PBC  and thus we are able to observe the looped spectrum in the complex plane.
 
 % \begin{figure*}[ht]
 % \label{figlast}
 % \end{figure*}

  \begin{figure*}[ht]
 \includegraphics[width=12.9cm]{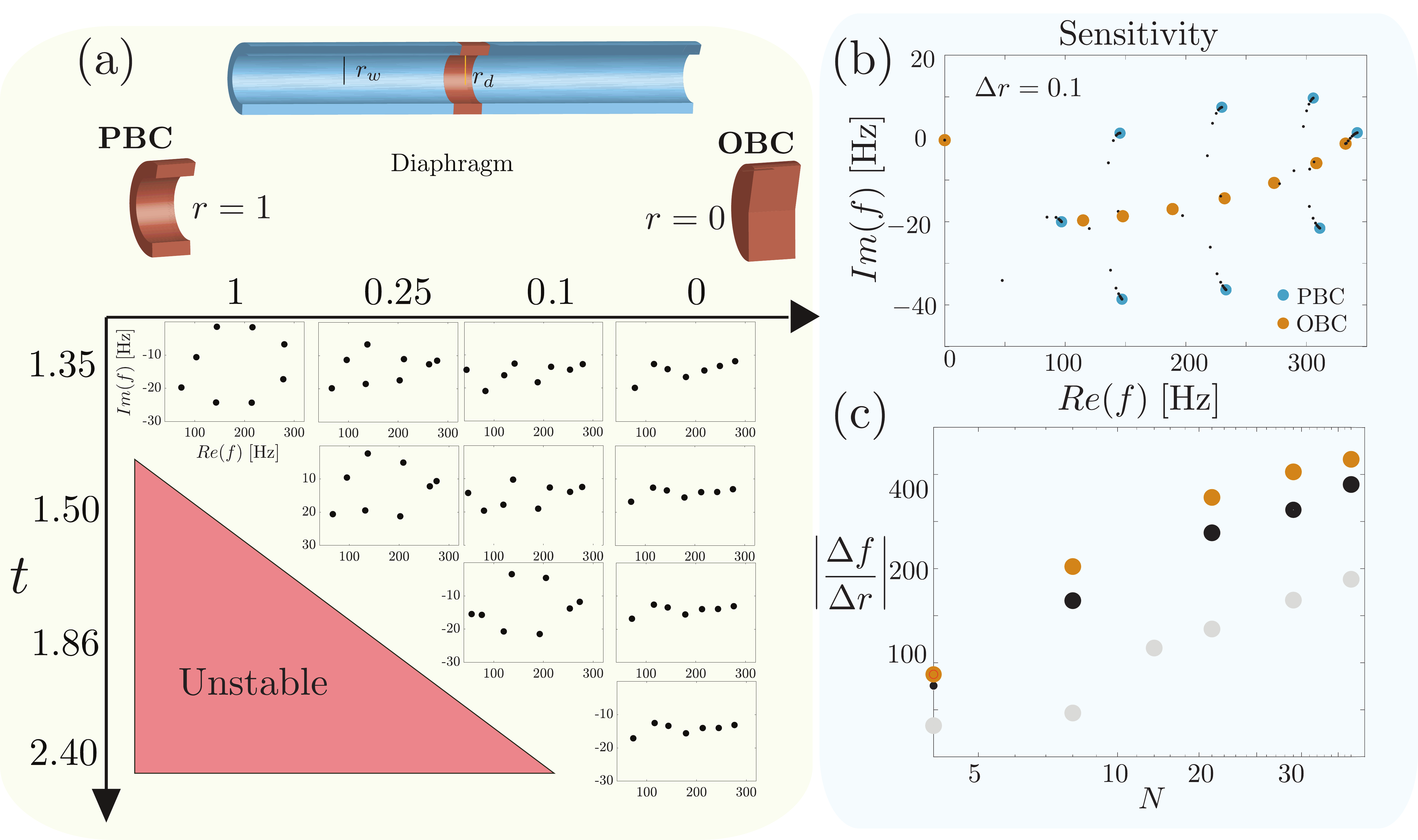}
 \caption{\textbf{Transition from PBC to OBC.} (a) A schematic of the diaphragm used in experiments inside the waveguide. The
eigenfrequencies obtained from experimental results for different values of the nonreciprocal hopping t and the ratio r. (b) The
transition from PBC to OBC obtain using the experimentally fitted transfer matrix. (c) The exponential sensitivity of absolute
value of the eigenfrequency as a function of the system size.}
 \label{figlast}
 \end{figure*}

Figure \ref{fig5}. shows the obtained eigenfrequencies of the acoustic problem in the complex plane for both the PBC and OBC configurations.
For the OBC, the reciprocal ($t=1$) and non-reciprocal ($t=1.35$) arrangements have seven acoustic modes and 
as predicted by the theory, they form an arc lying in the negative imaginary part of the complex plane. For the HN model the OBC spectrum always lies on the real axis for any value of $t$. However here we see that increasing the gain (thus $t$) the modes are pushed towards the real axis. This property, which is embedded in the proposed mapping, reveals the fact that adding gain to the system is able to better compensate losses. To further reveal the NHSE we measure the pressure field at different positions of the waveguide for the corresponding eigenfrequencies. An example of the mode shape for $t=3.7$ is shown in
Fig.\ref{figlast} where the energy is clearly localized predominantly on the right boundary ($j=8$).(b). Moreover, we plot the experimentally obtained inverse participation ratio 'IPR' $ \sum_k \vert p(x_k)\vert^2/(\sum_k \vert p(x_k)\vert)^2$ as a function of $t$ in Fig.\ref{figlast}.(c). This ratio quantifies the localization level of the eigenmodes, for instance a value of $\text{IPR}=1$ or $\text{IPR}=0$  indicates respectively  a total localization at one site or a full delocalization. As anticipated, the present results indicate that the increase in the hopping factor leads to a stronger localization of the eigenmodes on the right side of the system. 
% 
% \begin{figure}
%     \centering
%     \includegraphics[scale=0.4]{16_cells_t_2p4_theo.png}
%     \caption{$N=16$, $t=2.4$}
%     \label{fig:my_label}
% \end{figure}

In panel (d) Fig.\ref{figlast} we show the results for the PBC waveguide loaded with $N=8$. In the reciprocal case with $t=1$ it has 5 modes which among them three are degenerate with multiplicity 2 due the angular symmetry. On the other hand in the non-reciprocal case $t=1.35$, the degenerate modes split in pairs and the spectrum forms a closed loop in the complex plane indicated by the filled cirlcles in panel (d). An important aspect of the proposed acoustic system is that the inherent losses themselves allow for the obervation of the looped spectrum since all modes are stable. As expected, there is a maximum value of the gain after which some modes become unstable.

\subsection{ PBC to OBC and boundary sensitivity}

Another interesting aspect of the proposed system is that it allows to study experimentally 
the transition from PBC to OBC. This transition has been the subject of several studies since it gives insights on the sensitivity of the underlying spectrum under changes of the BCs. Experimentally it has only been observed in discrete lattices\cite{toponature} but not in a continuous wave system. Here, we achieve the transition in a rather natural way by adding a thin diaphragm of radius $r_d$ inside the looped waveguide of radius $r_w$ and progressively reducing the ratio $r=r_d/r_w$. In Fig.\ref{figlast}.(a) we plot the experimentally obtained spectrum in the complex plane for various values of $r$ between PBC ($r=1$) and OBC ($r=0$). The first row corresponds to a relatively small gain $t=1.35$, and the transition is explicitly demonstrated for all the ranges of the diaphragm radius (first line). As the radius decreases, the ellipse gradually shrinks until it transforms into an arc for the OBC.
This transition is clearly visible for various values of $t$ as shown in Fig.\ref{figlast} (a). In addition, as expected by the theory, by increasing the hopping factor for a fixed radius ratio (e.g. the column for $0.1$), the ellipse expands, which is a characteristic of the HN model. 

Interestingly, for higher values of $t$ the system becomes unstable before completely opening the diaphragm, since the mode with the largest imaginary part approaches the real axis. This is where the sensitivity of the system to boundary conditions is revealed. For the marginal case of  $t=2.4$ if we open a small hole ($10\%$ of the waveguide radius) to the stable OBC configuration abruptly becomes unstable.
In fact it was recently argued that the sensitivity of the HN with respect to the system size is exponential\cite{sensitivityPRL,sensitivityPRB,sensitivity2023}.

Due to the inherent instabilities and the limited number of cells used here, we cannot quantify this type of sensitivity directly from experiments. However, we do further investigate the sensitivity 
semi-analytically by using the experimentally obtained transfer matrix of the unit cell $M_{\mathrm{exp}}$. In particular, we use an analytical 1D model taking into account the waveguide, the cavity and the loudspeaker (see Fig.\ref{detT}) and fit it to the experimentally obtained elements of the transfer matrix. 
Then the  effect of a thin diaphragm, at low frequencies, can be approximated by assuming continuity of the acoustic flux and discontinuity of the acoustic pressure at the location of the diaphragm $x_d$ in the form \cite{KERGOMARD1987, Bolton} $$\left[\;p\;\right]_{x_d}=z(r,\omega) u(x_d).$$ The parameter $z$ includes both the resistive and reacting parts of the diaphragm (see Appendix). The corresponding transfer matrix of the defect can then be written as 
\begin{align} 
M_d=\begin{pmatrix}  1 &  -z\color{black}\\ 
0 & 1 \end{pmatrix}.
\label{Md}
\end{align}
For a periodic system with N-cells one can then calculate the solutions of $\mathrm{det}(M_dM_{\mathrm{exp}}^N-\mathds{1})=0$ to find the corresponding eigenfrequencies.
Figure \ref{figlast} (b) exhibits the eigenfrequencies of a system with $N=8$ and $t=1.5$ [corresponding to the second row of Fig.\ref{figlast} (a)]. Here we vary the ratio $r$ by increments of $0.1$.  What we observe is that 
with $r=0.1$ the eigenfrequencies have slightly shifted as observed in the experiments. Then for $r>0.2$ a large ellipse has been formed in complex plane indicating a strong change in the eigenfrequencies. To further quantify this sensitivity we have calculated the change in absolute value of frequency for the mode in the center of the ellipse for a change of the ratio $\Delta r=0.1$. The results for three different values of $t$ is shown in Figure \ref{figlast} (c) in a logarithmic scale. It is clear that the proposed acoustic system is indeed exponentially sensitive to its size.

\section{Conclusion}

As a conclusion, an exact map of the Hatano-Nelson model to one dimensional nonreciprocal continuous acoustic systems is presented. The mapping is achieved solely by using a transfer matrix approach, and can be applied to a plenitude of systems, provided that non-reciprocity can be implemented by the given device. The experimental results show the emergence of non-Hermitian skin effect once an asymmetric hopping is achieved, and by analyzing the complex frequency of the acoustic mode, the theoretical model is validated. Finally, while using diaphragms of different hole sizes, the transition from PBC to OBC, and the subsequent exponential sensitivity to the system is exhibited. Using the proposed method many other variances of the HN model can be constructed in continuous media, including various nonreciprocal topological models or  higher dimensional models  which profit from the interplay between topology and non-hermiticity.
%Among the most interesting ones, we can mention the addition of disorder in the network, the increase of the number of cells and a deeper study of the boundary sensitivity, studying the impact of non-neighboring hopping, or perhaps even implementing a similar approach (i.e. mapping through transfer matrices) for various NH models such as the SSH.

%
% \begin{figure}
%     \centering
%     \includegraphics[scale=0.4]{8_cells_t_2p4_theo.png}
%     \caption{$N=8$}
%     \label{fig:my_label}
% \end{figure}

\section{Acknowledgments}

V.A. acknowledges financial support from the NoHeNA project funded under the program Etoiles Montantes of the Region Pays de la Loire. V.A. Is supported by the EU H2020 ERC StG "NASA" Grant Agreement No. 101077954

\appendix
\section{Analytical transfer matrix}

Assuming a loudspeaker with a mechanical impedance $Z_m$ and force factor $B\ell$, is supplied by an electrical currant $i$, the equation of motion is given by,

$$Z_m v = (p_l-p_r)S_m + (B\ell) i$$

where the subscript $l$ and $r$ denotes the rear and forward face of cross section $S_m$.

Now, if the current is provided through an electroacoustic feedback with a static gain $G$, such as it satisfied the following equation $i=G p_l$, in addition to the conservation of flow, one gets the following transfer matrix,

\begin{align} 
M=\begin{pmatrix}  t & -iX-\beta\\ 
0 & 1 \end{pmatrix}
\label{Map}
\end{align}

where $t=1-\frac{GB\ell}{S_m^2}$, $\beta=\Re(Z_m/S_m^2)$, and $X=\Im(Z_m/S_m^2)$.

\section{Mapping energy to frequency}

\begin{align} 
t=(1-\alpha).
\end{align}

For the loudspeakers of the same kind as the ones used in Ref.\cite{penelet2021broadband}, the parameters in Eq.(\ref{Map})
are given by 
%with $g=\frac{c_0M_m}{Sm^2}$ and $k_0^2=(c_0M_mC_m)^{-1/2}$.
%In the other hand the losses and the reactance of the loudspeaker are present in the mapping Eq.(\ref{map}) with $B=\beta+i\mathcal{X}$ 
%where
\begin{align}
X=g\left(\frac{k^2-k_0^2}{k} \right),\quad \beta=\frac{S_W}{c_0\rho S_m^2}\left(R_m+\frac{(B\ell)^2}{Z_e}\right)
\end{align}
with $g=\frac{S_wc_0M_m}{c_0\rho Sm^2}$ and $k_0^2=(c_0^2M_mC_m)^{-1/2}$.
Note that $\beta$ quantifies the loudspeaker losses where the ratio between the loudspeaker membrane section $S_m$
and the waveguide section $S_w$ can be used as a tuning parameter.

\end{document}